\documentstyle[12pt,psfig]{article}
\begin{document}
\begin{titlepage}
\title{
\begin{flushright}
{\normalsize
SNUTP 96-129}
\end{flushright} \bigskip \bigskip
Chaos in the segments from Korean Traditional Singing and Western Singing}

\begin{center}
\author{Myeong-Hwa Lee\thanks{E-mail : mhwalee@gmc.snu.ac.kr}, 
Jeong-No Lee\thanks{E-mail : jnlee@gmc.snu.ac.kr}, 
Kwang-Sup Soh\thanks{E-mail : kssoh@phyb.snu.ac.kr}\\
{\small \it Department of Physics Education,
Seoul National University,
Seoul 151-742, Korea}}
\end{center}
\date{}
\maketitle

\begin{abstract}
We investigate the time series of the segments from a Korean traditional song 
``Gwansanyungma'' and
a western song ``La Mamma Morta'' using chaotic analysis techniques.
 It is found that the phase portrait in the reconstructed state space of the time 
series of the segment from the Korean traditional song has a more complex 
structure in comparison
with the segment from the western songs. 
The segment from the Korean traditional song has the correlation dimension 4.4
and two positive Lyapunov exponents which 
show that the dynamic related to the Korean traditional song
is a high dimensional hyperchaotic process. 
On the other hand, the segment from the western song with only one positive
Lyapunov exponent and the correlation dimension 2.5 exhibits 
low dimensional chaotic behavior.
\end{abstract}
\nonumber
\end{titlepage}

\newpage
\nonumber
\section{Introduction}
\hspace{0.5cm}
Until recently it has been thought that complicated phenomena result from complicated
dynamical systems with many degrees of freedom and  they have been 
usually considered as stochastic
processes. 
But it turned out that even simple systems with few degrees of freedom
can exhibit extremely irregular unpredictable behavior when the systems
are described by nonlinear deterministic equations.
The most typical example is chaos which shows extremely
sensitive dependence on initial conditions.

Several authors have analyzed speech data using techniques 
for studying chaotic
dynamics\cite{speech}. Despite the apparently complicated mechanism of vocal production, 
which involves various aspects such as physiological activities and 
nonlinear fluid mechanics,
 the concept of deterministic chaos implies the possibility 
that the complex irregularities may be
generated by deterministic nonlinear dynamics with only a few state variables.
It has been reported that human speech sounds exhibit bifurcations and
chaos\cite{bifur}, for example, in fricative consonants\cite{conso}, vowels\cite{vowel},
newborn infant cries\cite{newborn}, various phoneme types\cite{nonlin} 
and Korean traditional
singing\cite{newp}.
In this paper we analyze the irregularities and the dynamics in the time series 
of the segments from a Korean traditional 
song and a western song. 
In order to check our method of analysis
a single note ``Si'' is
also analysed.
In section 2, our data acquisition from the Korean traditional song,
the western song  and the note ``Si' is explained.
In section 3, the power spectra of the time series are given and
in section 4, the phase portraits which give geometrical information included in the time series
are constructed by the delay vector method. In section 5,
 the correlation dimension of the attractors in the reconstructed space are calculated to characterize
properties of the attractor.
In section 6, Lyapunov exponents are estimated.
The above mentioned three songs are compared in sections three through six.
In the last section brief discussion is given.
 
\section{Data Acquisition}
\hspace{0.5cm}
In each section, for the sake of comparison, segments of a Korean traditional song,
segments of a western
song and the note ``Si'' are analyzed at the same time.
The song used as Korean traditional singing, ``Gwansanyungma'' is a SeoDo folk song
which has been sung in the northwestern provinces of Korea. 
It includes various kinds of vocalization of
SeoDo folk song, especially the peculiar vibration. 
The song we used was sung in 1994 by Pok-Ny\v{o} O 
who was born in P'y\v{o}ngyang in 1913 and 
was designated as one of the human cultural assets in 1971.
Its length is about 220 seconds and its words and phonetic transcription
using single-symbol ARPAbet are given in Table \ref{tab}.
As seen in Table \ref{tab}, they are divided into 19 sections to identify which
sections show chaotic behavior.
We calculated the power spectrum for each section. 
After getting insight about the characteristics of each section from the
power spectrum, we choose the 8th section for obtaining the 
phase portrait, correlation dimension and Lyapunov exponent 
because the 8th section showed the most broadband power spectrum 
and was thought to
represent the characteristics of SeoDo folk song most properly.
Actually only the latter part of the 8th section is used, whose length
is about 1.6 seconds and
its phoneme is described by the vibrating vowel /a/. 
As a western song,``La Mamma Morta'' performed by Maria Callas
is used. Especially the segments which don't have background music are used 
to facilitate the analysis. 
Its length is about 1 second and the phoneme is also /a/.
The single note ``Si'' of the ``Do-re-mi song'' in the movie `Sound of music' 
is used to check our methods of analysis. Its length is about 0.5 second.
The consonant /s/ is uttered only for a very short time, so the actually
recorded and analysed sound is /i/.

All the data are recorded in the form of wave files with PCM (Pulse Code
Modulation) mode using `Sound Blaster
AWE32' which is the trademark of Creative Technology Ltd.
It is performed with sampling rate 22050 Hz and 16 bits
except the single note ``Si'' with 44100 Hz owing to its short length.
Programs to analyze the recorded wave files run on an IBM PC.

\section{Power spectrum}
\hspace{0.5cm}
The Fourier analysis of time series is useful for obtaining, among other things,
the frequency components and the power distribution as a function of
frequency.
In general the power spectrum is a very good tool for the visualization of
periodic and quasi-periodic phenomena, and their separation from irregular time 
evolutions which are characterized by broadband power spectra\cite{sa}. 
The broadband regions
in power spectra can arise from stochastic or deterministic processes, but 
the decay in the spectral power at large $\omega$ is different for the two cases\cite{greenside}
\cite{sigeti}. However, that is not really a good indicator to 
identify chaotic time series. So we need to estimate correlation dimension and 
Lyapunov exponents,etc.
We use the FFT to compute a power spectrum. 

\begin{figure}[]
\centerline{\psfig{file=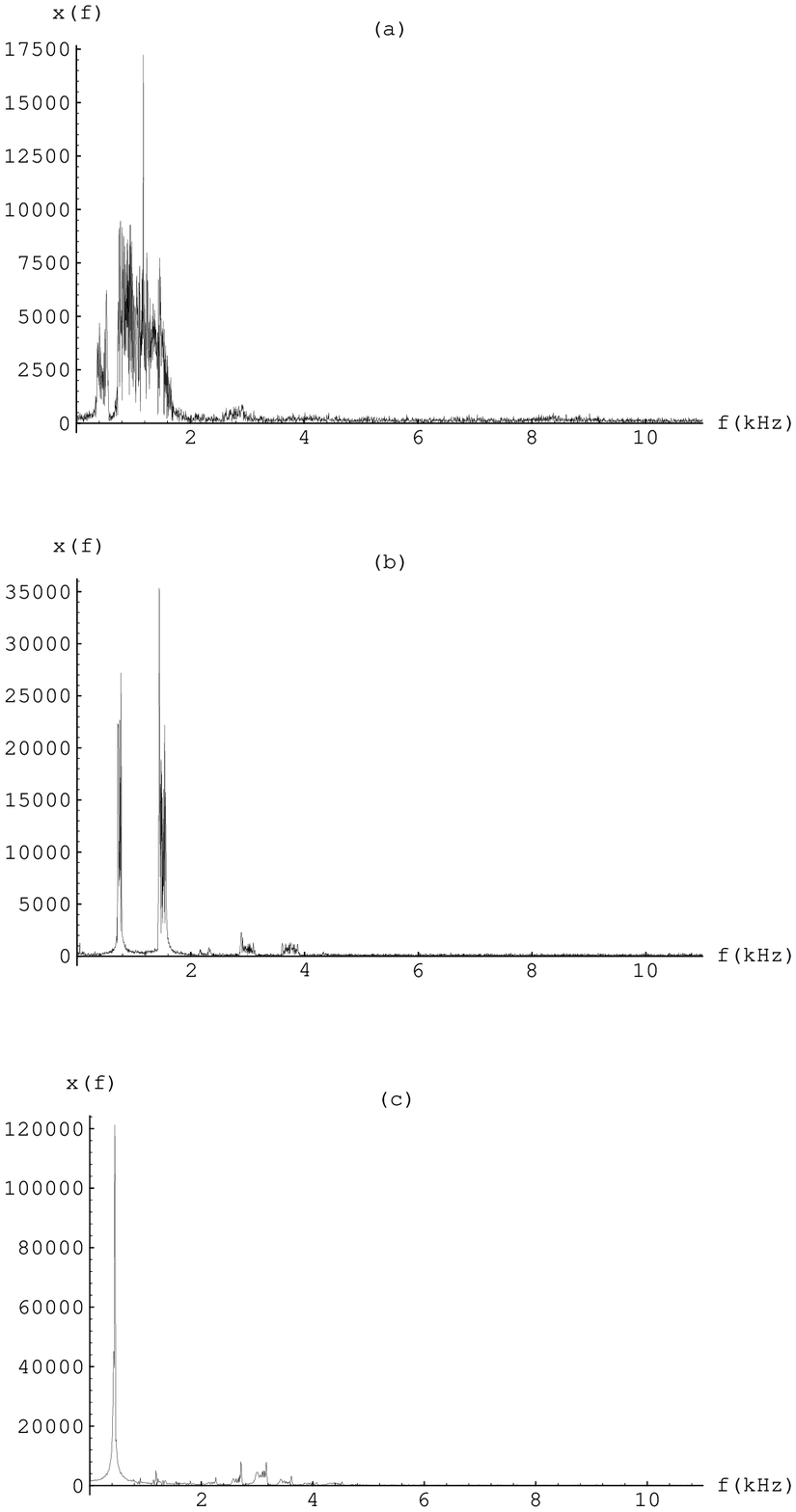,width=130mm,angle=0}}
\caption{(a) Fourier Transform of SeoDo folk song shows
the very broadened spectrum  which is one of the
characteristics of chaotic system.~
(b) Fourier Transform of western song shows
a little broadened spectrum.~
(c) Fourier Transform of ``Si'' shows
$\delta$-function which is the characteristics of
the periodic system.}
\label{fft}
\end{figure}

In the case of Korean traditional song it is obtained for each section
divided in order to identify whether each section shows chaotic behavior or not.
The sections with broad power spectrum are marked with ``$\ast$'' in Table \ref{tab}.
Figure \ref{fft} shows the power spectrum obtained from the time series of 
the Korean traditional song,
western song
and a single note ``Si''.
In the case of Korean traditional song it is that of the most typical case
with broad power spectrum(8th section). 
We can see the broadband regions in the power spectrum of the Korean traditional
song which is a characteristics of chaotic systems. The power spectrum of 
the western song shows  
less broadband power
 in comparison with the Korean traditional song and we can see $\delta$-function 
like peak
in power spectrum of a single note ``Si''. 
As the data to investigate phase portrait, 
correlation dimension and
Lyapunov exponents for the Korean traditional song, 
we choose the 8th section which
shows the most broadband power spectrum. 

\section{Phase portraits}
\hspace{0.5cm}
Several variables are usually required to fully describe the state of a
system. Thus, in order to construct a dynamical model for an experimental time series, 
we must
first reconstruct a state space\cite{svd}.

A simple way to create a state representation is to form vectors of
the $n$ consecutive samples of the time series which are delay vectors ${\bf X}_{t}$. The delay vector,
${\bf X}_t$, at time $t$ is defined by:
\begin{equation}
{\bf X}_t\equiv (x_t, x_{t+T},\ldots,x_{t+(n-1)T})
\end{equation}
where $n$, $T$ are called the {\it embedding dimension}, {\it delay time}, respectively and $\{x_t\}$
stands for the time series.
This choice of state representation is commonly known as {\it delay vector method}. 
Takens\cite{takens} proved that if $d$ is the dimension of a
manifold containing the attractor, in order to yield a topological
 equivalent phase portrait, leaving the dynamical parameters invariant, it is sufficient to choose $n$ such that:
\begin{equation}
n\geq 2d+1
\end{equation}
The choice of optimal delay time $T$ is an important and largely unresolved problem.
Even though the choice of delay time $T$
is arbitrary for an infinite amount of noise-free data\cite{takens}, in the case of real laboratory data 
a good choice of $T$ is essential for geometrical and numerical analysis of a phase portrait. 
As $T \rightarrow 0$, the trajectory approaches the identity 
line. When $T$ is larger than the optimal value, the correlation between the data points is destroyed.

A commonly used rule of thumb is to set $T$ to be the time lag required for the autocorrelation
function to become negative, or alternatively, the time lag required for the autocorrelation to
decrease by a factor of $e$ (from its value at time lag $0$ ). Another approach, that of Fraser
and Swinney\cite{fraser} uses the ``mutual information'', which measures a more general dependence of two variables.
Schuster and Liebert\cite{liebert} took the first minimum of the logarithm of the 
generalized correlation integral as a criterion for a proper choice of the delay time. 
In choosing the optimal delay time, our scheme is as follows: At first we obtain 3-dimensional 
projections of phase portraits with different delay times and then
choose delay time $T$ at which value overfolding or the straight line do not appear
in the 3-dimensional projections as an optimal delay time.
Once we reconstruct a $n$-dimensional space in this way, we can obtain a good
representation of the attractor and also obtain an appropriate correlation dimension 
with this delay time. 
We also obtain the phase portraits for delay time at which value the autocorrelation
decreases to $e^{-1}$ of its initial value and 
the result is nearly the same as we use the delay time 
obtained
with 3-dimensional projection.
 
\begin{figure}[b]
\centerline{\psfig{file=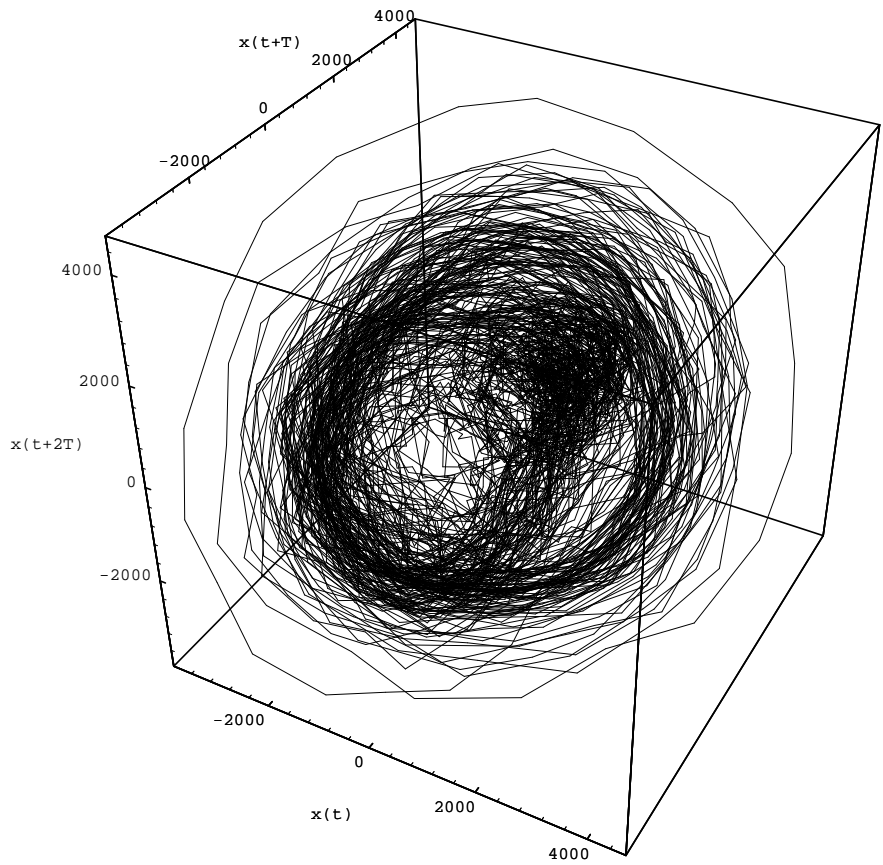,width=130mm,angle=0}}
\caption{3-dimensional phase portrait of SeoDo folk song
wrinkles and occupies an extended region.}
\label{ppe}
\end{figure}

\begin{figure}[]
\centerline{\psfig{file=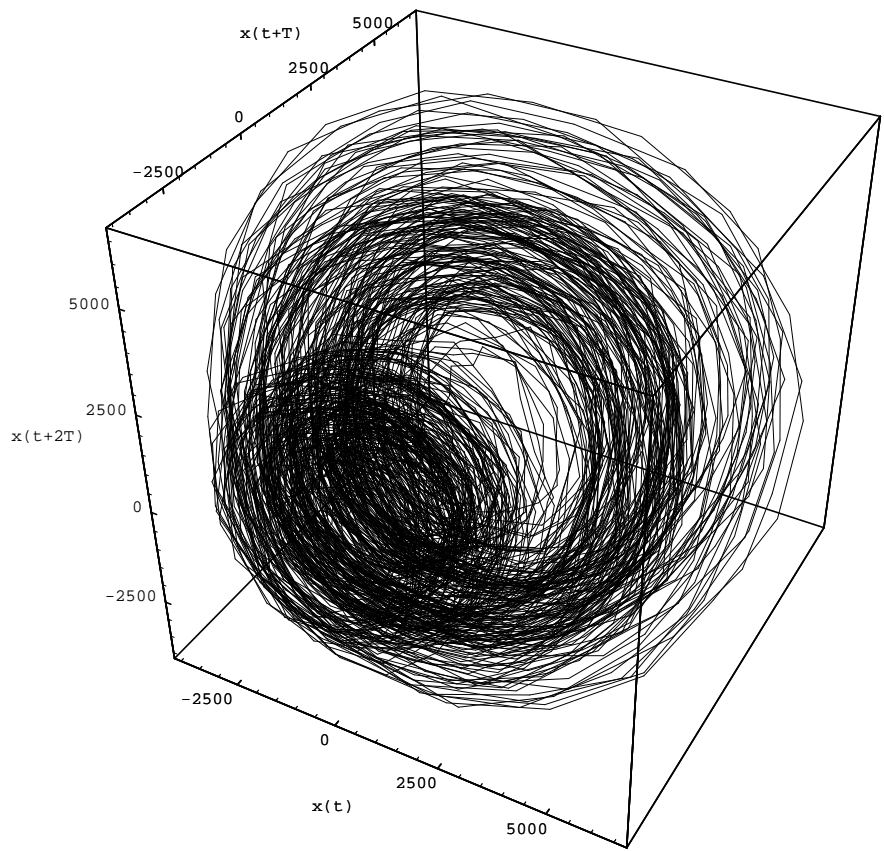,width=130mm,angle=0}}
\caption{3-dimensional phase portrait of western song also
wrinkles and occupies an extended region.}
\label{ppw}
\end{figure}

\begin{figure}[h]
\centerline{\psfig{file=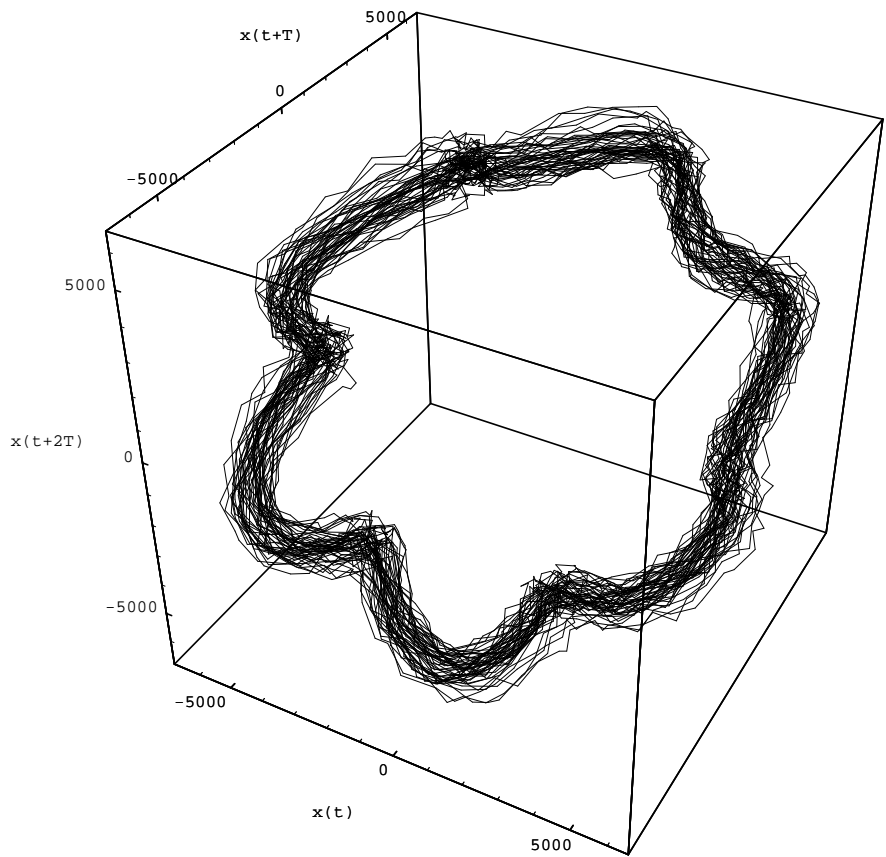,width=130mm,angle=0}}
\caption{3-dimensional phase portrait of ``Si''
exhibits a limit cycle.}
\label{pps}
\end{figure}

Figures \ref{ppe}-\ref{pps} show the 3-dimensional phase portraits of 
the segments from the Korean 
traditional song, 
the western song
and the single note ``Si'' respectively. 
We can see that the phase portrait of the segments from the Korean traditional song
wrinkles and occupies an extended region.  
This does not necessarily imply that the behavior
is stochastic, but rather that the dimension of the strange attractor is too
large to be determined by visual inspection of the attractor. For these
attractors the dimension must be calculated quantitatively.
The phase portrait of the segment from the western song also shows 
complicated and tangled structure.
Comparing these two phase portraits
we can infer that the attractor of the segment from the Korean traditional song 
will have a higher value
of dimension than that of 
the segment from the western song because 
it has a more complex structure in comparison with the western song.
The phase portrait of single note ``Si'' is supposed to represent a limit cycle
but has thickness due to noises.
It is essentially a periodic system and accords with the result of 
the power spectrum.

\section{Correlation dimension}
\label{correlation}
\hspace{0.5cm}
The dimension of a phase portrait is clearly the first level of knowledge necessary
to characterize its properties and also gives us a lower bound on the number 
of essential variables needed
to model the dynamics.
We calculate the correlation dimension which is relatively easy to compute.
It can be calculated by measuring the correlation integral, 
which represents the spatial correlation of the points on the attractor,
defined by\cite{grassberger}
\begin{equation}
C(r) =\lim_{N\to\infty}\frac{2}{N(N-1)}
      \left\{ \begin {array} {c}
              {\rm the~number~of~pairs~ of~ points},~({\bf X}_i,{\bf X}_j),\\
              {\rm on~ the~phase~portrait~ with~ separation~} <~ r
              \end{array}
       \right\},
\end{equation}
then one observes that as $r\to 0$,
\begin{equation}
C(r)\propto r^\nu,
\end{equation}
giving the working definition of the correlation dimension, $\nu$.
But it is known that for limited data sets with high autocorrelation
the correlation integral displays an anomalous shoulder which hinders
good estimates of dimension\cite{theiler}.
It is because the distances between these pairs of points 
which are not well separated in time do not really reflect 
the geometrical properties of the attractor.
So we use the modified correlation integral $C(W,N,r)$,
which discards the pairs of points closer together than $W$ 
in time\cite{theiler2},
\begin{equation}
C(W,N,r) =\frac{2}{(N+1-W)(N-W)}\sum_{n=W}^{N-1}\sum_{i=0}^{N-1-n}H(r-\parallel
X_i-X_{i+n}\parallel),
\end{equation}
which gives the standard algorithm when $W=1$.
In this calculation the autocorrelation time is used as a value of $W$. 

Plotting the value of $C(W,N,r)$ versus $r$ on a log-log plot gives a slope 
$\nu$ for small $r$.
From this plotting of $\log{C(W,N,r)}$ versus $\log{r}$, one can distinguish low dimensional
deterministic chaotic signals from stochastic noise.
 For the deterministic signal, if one increases the embedding
dimension $n$, the slope of $\log C(W,N,r)$ versus $\log r$ at first also 
increases but settles
at a value of $\nu$ and becomes independent of the embedding dimension. 
For random noise, the 
slope of $\log C(W,N,r)$ versus $\log r$ increases indefinitely as the embedding 
dimension $n$ is increased.

\begin{figure}[h]
\centerline{\psfig{file=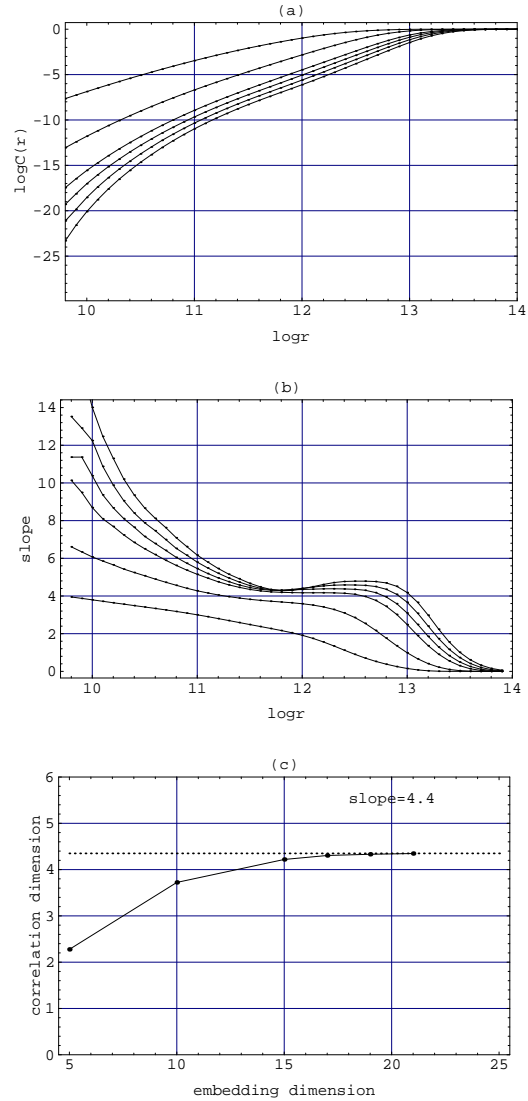,width=130mm,angle=0}}
\caption{Correlation integral and correlation
dimension for SeoDo folk song;
(a) The plot of $\log_{2.0}C(W,N,r)$ versus $\log_{2.0}r$ for
values of r
varying from $2.0^{9.8}$ to $2.0^{14}$ with embedding
dimension 5,10,15,17,19,21 (top to bottom).
(b) The plot of local slopes versus $\log_{2.0}r$ showing
converging scaling region.
(c) Correlation dimension $\nu$ is saturated at a value
of about 4.4
as the embedding dimension is increased.}
\label{ce}
\end{figure}

\begin{figure}[h]
\centerline{\psfig{file=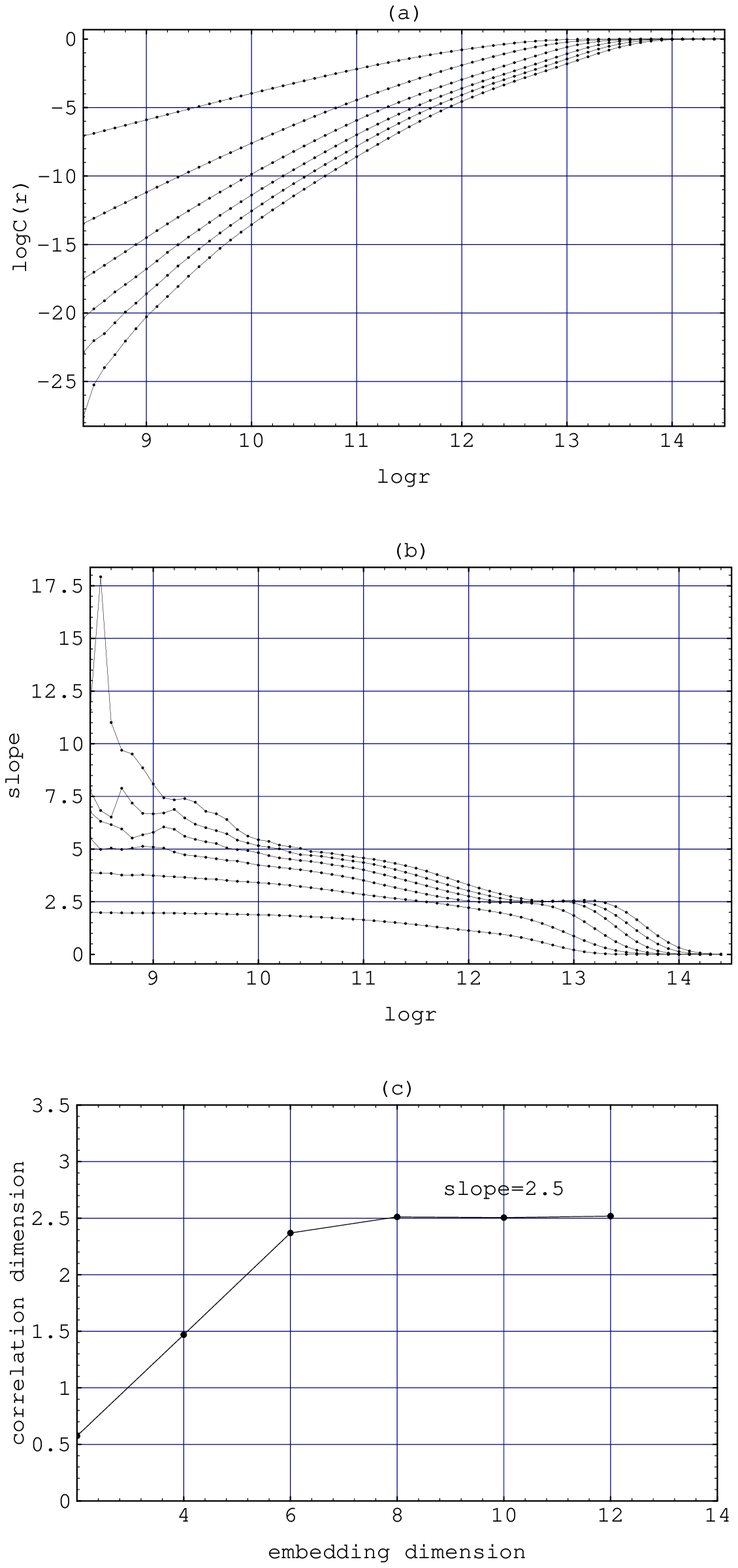,width=130mm,angle=0}}
\caption{Correlation integral and
correlation dimension for western song;
(a) The plot of $\log_{2.0}C(W,N,r)$ versus $\log_{2.0}r$ for
values of r
varying from $2.0^{8.4}$ to $2.0^{14.5}$ with embedding
dimension 2,4,6,8,10,12 (top to bottom).
(b) The plot of local slopes versus $\log_{2.0}r$ showing
converging scaling region.
(c) Correlation dimension $\nu$ is saturated at a value
of about 2.5
as the embedding dimension is increased.}
\label{cw}
\end{figure}

\begin{figure}[h]
\centerline{\psfig{file=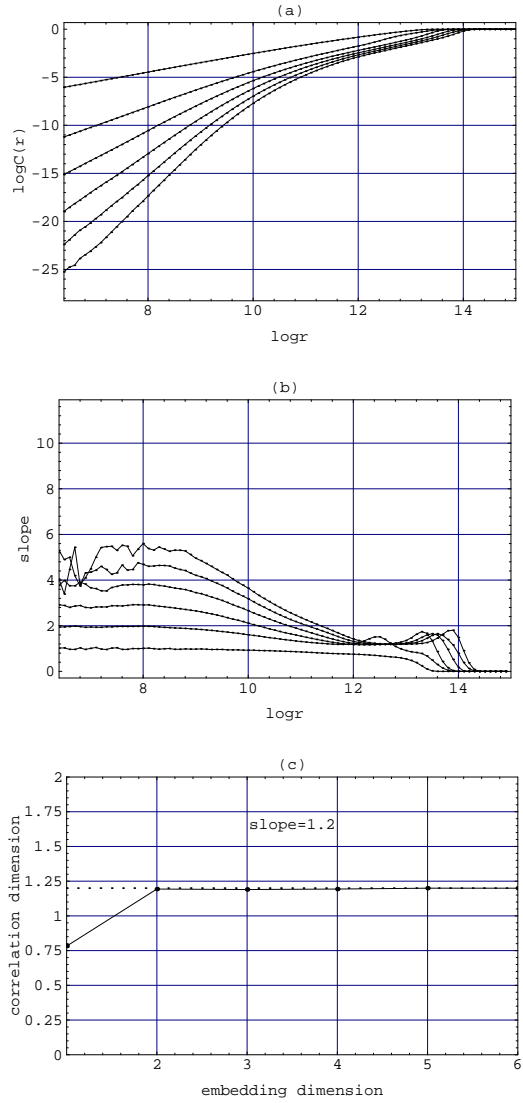,width=130mm,angle=0}}
\caption{Correlation integral and correlation dimension
for ``Si'';
(a) The plot of $\log_{2.0}C(W,N,r)$ versus $\log_{2.0}r$ for values of r
varying from $2.0^{6.3}$ to $2.0^{14.5}$ with embedding
dimension 1,2,3,4,5,6 (top to bottom).
(b) The plot of local slopes versus $\log_{2.0}r$ showing
converging scaling region.
(c) Correlation dimension $\nu$ is saturated at a value
of about 1.2
as the embedding dimension is increased.} 
\label{cs}
\end{figure}

The correlation dimension
is measured using $3.5\times 10^4$ data points for time series
of the segment from the Korean traditional song.
The slope $\nu$ of $\log C(W,N,r)$ versus $\log r$ plot of 
the segment from the Korean traditional
song is saturated at a value
 of about 4.4 independent of the embedding dimension, which we assume 
to be the correlation
 dimension of the attractor (Figure \ref{ce}).
The slope $\nu$ of $\log C(W,N,r)$ versus $\log r$ plot of 
the segment from the western
song measured using $2.0\times 10^4$ data points 
is saturated at a value of about 2.5 independent of the embedding 
dimension (Figure \ref{cw}).
The correlation dimension of the attractor obtained from a single note ``Si'' 
with $2.0 \times 10^4$ data points is about 1.2 (Figure \ref{cs}).
It is thought that a noise causes the correlation dimension to be saturated 
at a higher value
than the expected value 1.0. 
In these all cases we calculated the correlation dimensions for
several reasonable values of delay time, which are obtained from 
phase portrait and autocorrelation.
The correlation dimension itself was almost the same except that
when delay time is such that the autocorrelation of time series
decreases to $e^{-1}$ of its initial value
the correlation dimension is saturated at lower embedding dimension.  

\section{Lyapunov exponents}
\label{lyapunov}
\hspace{0.5cm}
Lyapunov exponents are one of the most important quantities used to
distinguish chaotic from nonchaotic behavior. 
When a dynamic is chaotic, positive
Lyapunov exponents occur that quantify the rate of separation of 
neighboring (initial) states
and give the time duration where predictions are possible. 
For such experimental works as our case,
 methods for computing the Lyapunov exponents of a given time series are 
of great importance.
The different algorithms that have been proposed during the last decade 
for this purpose
are all based on the reconstruction of the corresponding attractor 
by means of delay vector method
with suitable embedding dimension. Most of the methods are based on approximations of
the unknown flow governing the dynamics on the reconstructed attractor 
in embedding space.
From the sequence of Jacobians of this flow at the state points along 
the reconstructed orbit,
the Lyapunov exponents can be computed by standard methods\cite{lypo85}\cite{Eckmann}.
The crucial point of this approach is the accuracy of the approximation of the Jacobian
of the unknown flow. The most often used algorithms to achieve this goal are based on 
linear approximations. In this paper, we used Eckmann \& Ruelle
algorithm\cite{Eckmann}.

In the Eckmann $\&$ Ruelle algorithm, it is very important to choose 
appropriate values of 
the embedding dimension 
$n$ and range $r$. We choose the range, $r$, which is about 3-5$\%$ 
of the horizontal 
extent of the analysed 
attractor\cite{lypo85}. 
The delay vector to calculate Lyapunov exponents is obtained by the
same way for the same data as in the case of 
correlation dimension. 
We estimate the Lyapunov spectra at various embedding dimensions 
and then identify the true Lyapunov exponents and the spurious Lyapunov exponents 
on the point of view that true Lyapunov exponents do not vary much over 
a range of $n$, while
the spurious ones wander with $n$\cite{kruel}.
We also calculate the Lyapunov exponents for the different parameters(delay time,
range, e.t.c.) and the result is such that the sign of the largest
Lyapunov exponents does not change.

\begin{figure}[h]
\centerline{\psfig{file=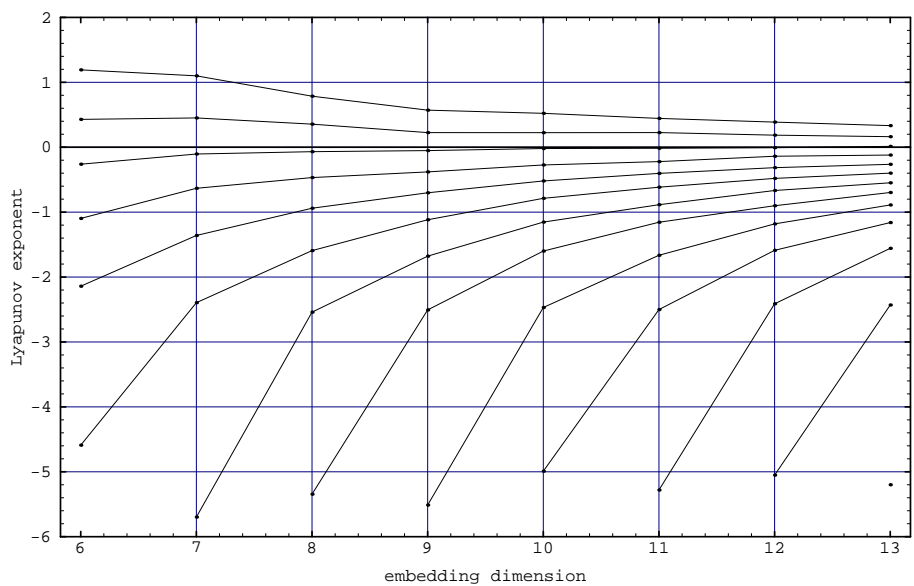,width=130mm,angle=0}}
\caption{SeoDo folk song has two positive Lyapunov exponents
under varying embedding
dimension $6-13$.}
\label{le}
\end{figure}

\begin{figure}[h]
\centerline{\psfig{file=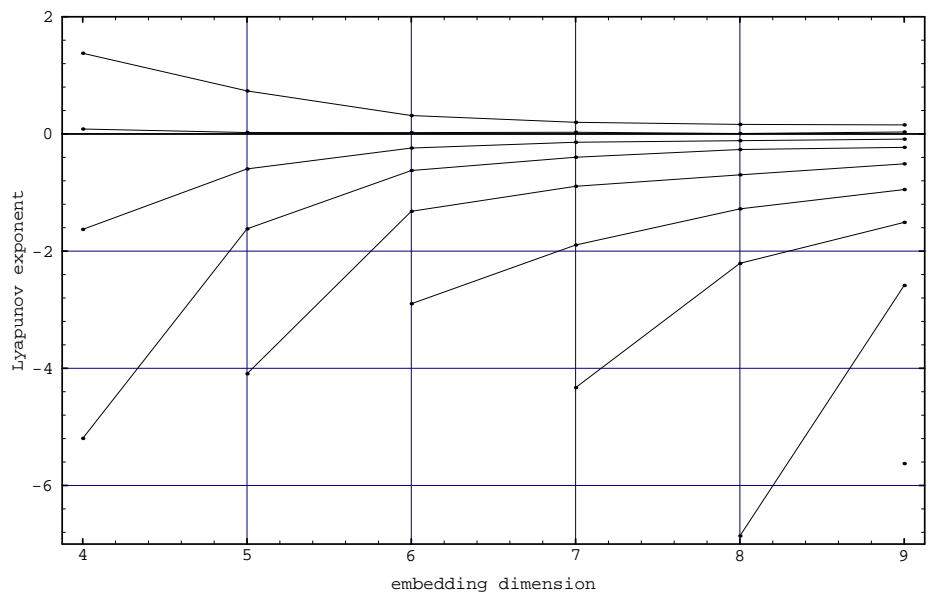,width=130mm,angle=0}}
\caption{Western song has one positive Lyapunov exponent
under varying embedding dimension $4-9$.}
\label{lw}
\end{figure}

\begin{figure}[h]
\centerline{\psfig{file=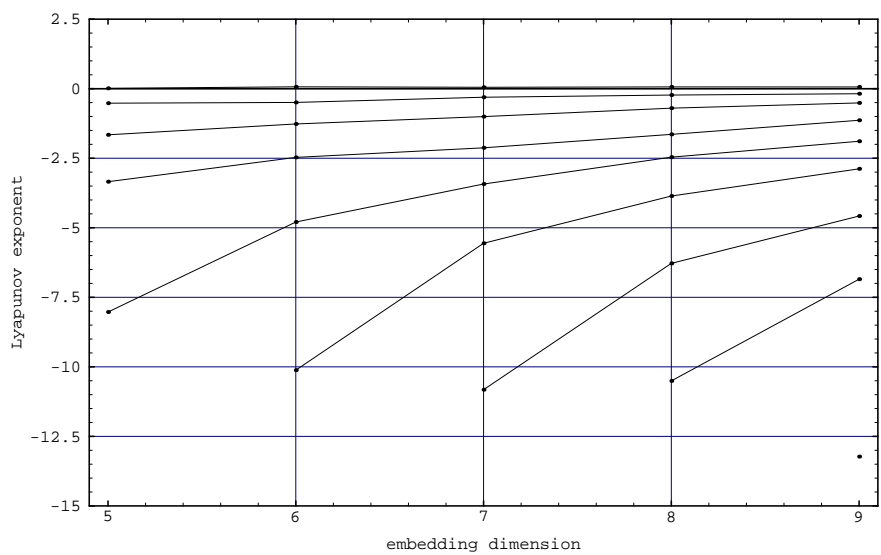,width=130mm,angle=0}}
\caption{Lyapunov exponents for ``Si''
under varying embedding dimension $5-9$.
It has one zero value and all negative values.}
\label{ls}
\end{figure}

Figures \ref{le}-\ref{ls} show the Lyapunov exponents with $n$ increased using 
the time series of the segments from
the Korean traditional song, the western song and ``Si''. 
We obtain two positive Lyapunov exponents in time series 
of the segment from the Korean 
traditional song, which
indicate that the attractor is a hyperchaotic attractor with two directions of 
expansion.
It is known that it is possible to find hyperchaotic attractors 
with two or more positive Lyapunov exponents only 
in higher (at least four) dimensional systems\cite{hyperchaos}, 
which accords with our result that the correlation 
dimension of the Korean traditional song is about 4.4. 
From the segment of the western song 
we obtain one positive 
Lyapunov exponent which means that the dynamic is a chaotic process. 
The Lyapunov dimension for the time series of segments from the Korean 
traditional song and western song is calculated by the 
Kaplan-Yorke formulae\cite{ky}.
In the case of the segment from the Korean traditional song we obtain 4.5-5.3 
as the Lyapunov dimension for the embedding dimension from 9 to 13, and 
3.1-3.4 
with the embedding dimension from 6 to 9
for the segment from the western song.
The result turns out that the Lyapunov dimensions in these two cases 
are a little higher than the 
correlation dimensions, but it is known that the Lyapunov dimension
forms an upper limit for the information dimension as far as higher-dimensional
systems are concerned\cite{kyd}.
As we expect,  
Lyapunov exponents from the single note ``Si''
is $(0,-,-,\cdots)$, which exhibit characteristics of a limit cycle.

\section{Discussion}
\hspace{0.5cm}
In this paper, we examine the music signals from the Korean traditional 
song ``Gwansanyungma'', 
the western song ``La Mamma Morta'' and the single
note ``Si'' through power spectra, phase portraits, 
the correlation dimension and Lyapunov 
exponents. From the time series of 
the segments from the Korean traditional song, 
we obtain a broadband power spectrum 
and the phase portrait with a complex and wrinkled structure 
in the reconstructed phase space. 
From the result of estimating the correlation dimension and the Lyapunov 
exponents, we can conclude 
that the underlying dynamics of the segment from 
the Korean traditional song is a high dimensional 
($\nu = 4.4$) hyperchaotic system
which has two directions diverging exponentially. 
The time series of the segment from the western song 
gives a power spectrum with a little 
broadband power and 
its phase portrait occupies an extended region.
Measurement of correlation dimension of it revealed that 
its trajectory lies 
on an attractor with dimension
about 2.5. The result that one of Lyapunov exponents of the segment
from the western song 
is a positive value shows that it
is a chaotic system of which nearby trajectories exponentially 
diverge on the average.
The time series of the single note ``Si'' which is chosen to 
confirm our analysis 
methods shows 
that it is from a periodic system as we expect, that is,
it has $\delta$-function shaped power spectrum,
a limit cycle blurred by noise in the reconstructed phase space, 
the correlation dimension of about 1.2
and the largest Lyapunov exponent zero.

In conclusion, we can observe hyperchaotic behavior in the 
segment from the Korean traditional song
and chaotic property in the segment from the
western song. It gives some hints for modelling certain aspects
of the sound production of the vocal system, especially approaching 
the vocalization methods. 
Large dimensionality of the segment from the Korean traditional song 
in comparison 
with the segment from the western song means that
its system is more complex and has more degrees of freedom. 
It is thought that it results from the peculiar vocalization of the 
Korean traditional song, 
especially vibrating techniques of SeoDo folk song. 

We did not investigate the whole parts but only the segments from
the Korean traditional song and the western song, which are
thought to reflect adequately the characteristics of the Korean traditional 
song and the western song.
Further study is needed to generalize above results 
by investigating more data sets and relate the method of vocalization 
with chaos more precisely. 

\vspace{1cm}
\begin{center}
{\large \bf ACKNOWLEDGMENTS}
\end{center}
\hspace{0.5cm}
This work was supported in parts by the Center for Theoretical
Physics, Seoul National University, and by the Basic Science
Research Institute, Ministry of Education under Project 
No. BSRI-96-2418.

\newpage
\begin{table}[h]
\caption{\label{tab} Divided Sections of the SeoDo Folk Song
``Gwansanyungma''}
\begin{center}
(The sections with broad power spectrum are marked with
``$\ast$''.)
\end{center}
\begin{center}
\vspace{4mm}
\begin{tabular}{|l|c|c|r|} \hline 
No & Words  & Phonetic transcription & Time  \\ \hline 
1 & ch'ugangi & /CUgaGi/  & 35 s             \\ \hline 
2 & i $\sim$  & /i/  & 10.0 s              \\ \hline
3 & i $\sim$  & /i/  & 14.5 s              \\ \hline
4 ($\ast$) & ch\v{o}gma  & /Jxma/ & 10 s    \\ \hline
5 ($\ast$) & ak          & /ak/  & 9.0 s    \\ \hline
6 ($\ast$) & \v{o}ryong  & /xrycG/ & 6.0 s      \\ \hline
7 & naeng  & /nEG/     & 12.5 s              \\ \hline
8 ($\ast$) & ha $\sim$ & /ha/  & 11.5 s     \\ \hline
9 ($\ast$) & ni       & /ni/   & 8.5 s      \\ \hline
10 ($\ast$) & i $\sim$  & /i/  & 9.5 s      \\ \hline
11 ($\ast$) & hi $\sim$ & /hi/  & 7.5 s      \\ \hline
12 ($\ast$) & injae     & /inJE/  & 3.0 s      \\ \hline
13 & aei    & /Ei/    & 9.0 s               \\ \hline
14 ($\ast$) & i $\sim$ & /i/  & 9.5 s      \\ \hline
15 ($\ast$) & s\v{o}  & /sx/   & 7.0 s      \\ \hline
16 & p'u     & /pu/   & 13.0 s              \\ \hline
17 & u $\sim$  & /u/  & 8.0 s               \\ \hline
18 ($\ast$) & $\sim$ ung & /uG/ & 13.0 s      \\ \hline
19 & chungs\v{o}nrur\v{u}l & /JUGsxNrUrL/ & 15.0 s  \\ \hline 
\end{tabular}
\end{center}
\begin{center}
(``$\sim$'' means that the peculiar vibration of SeoDo folk
song is used.)
\end{center}

\end{table}

\newpage
\pagestyle{empty}

\end{document}